\documentclass[a4paper,11pt]{article}
\usepackage[T1]{fontenc}
\usepackage{pos}
\renewcommand{\logo}{\relax} 
\renewcommand{\posurl}{\relax} 
\usepackage{amsmath}
\usepackage{mathtools}
\usepackage{tikz}
\usepackage{pgfplots}
\pgfplotsset{compat=1.5}
\usepackage{chngcntr} 
\counterwithin{figure}{section}    
\counterwithin{table}{section}
\counterwithin{equation}{section}
\usepackage{color}
\usepackage{siunitx}
\usepackage{multicol}
\usepackage{sectsty}
\usepackage{changepage}
\usepackage{natbib}
\DeclareSIUnit\yr{yr}
\DeclareSIUnit\ns{ns}
\DeclareSIUnit\MW{MW}
\DeclareSIUnit\min{min}
\DeclareSIUnit\h{h}
\DeclareSIUnit\day{d}
\DeclareSIUnit\d{d}
\DeclareSIUnit\PeV{PeV}
\DeclareSIUnit\TeV{TeV}
\DeclareSIUnit\EeV{EeV}
\DeclareSIUnit\MB{MB}
\DeclareSIUnit\GB{GB}
\DeclareSIUnit\TB{TB}
\DeclareSIUnit\pc{pc}
\DeclareSIUnit\kpc{kpc}
\DeclareSIUnit\Mpc{Mpc}
\DeclareSIUnit\inch{inch}
\DeclareSIUnit\kt{kt}
\DeclareSIUnit\atm{atm}
\DeclareSIUnit\PE{PE}
\DeclareSIUnit\HP{HP}
\DeclareSIUnit\pb{pb}
\DeclareSIUnit\permille{\text{\textperthousand}}

\definecolor  {blue1}{rgb} { 0.50, 0.50, 0.90 }
\definecolor  {blue2}{rgb} { 0.90, 0.90, 0.95 }
\definecolor  {blue3}{rgb} { 0.20, 0.20, 0.60 }

\definecolor   {red1}{rgb} { 1.00, 0.20, 0.20 }
\definecolor   {red2}{rgb} { 1.00, 0.90, 0.90 }
\definecolor   {red3}{rgb} { 0.80, 0.00, 0.00 }

\definecolor {green3}{rgb} { 0.00, 0.50, 0.00 }

\definecolor {yellow1}{rgb}{ 0.95, 0.70, 0.00 }
\definecolor {yellow2}{rgb}{ 0.80, 0.50, 0.00 }

\definecolor {magenta1}{rgb} { 1.00, 0.50, 1.00 }

\definecolor {cyan1}{rgb} { 0.05, 0.50, 0.40 }
\definecolor {cyan2}{rgb} { 0.55, 0.80, 0.70 }
\definecolor {cyan3}{rgb} { 0.20, 0.70, 0.70 }

\newcommand{\bfr}[1]{\textbf{\color{red}#1}}
\newcommand{\tit}[1]{\text{\textit{#1}}}
\newcommand{\hide}[1]{}
\newcommand{\remind}[1]{\bfr{[#1]}}

\newcommand{\dcp}{$\delta_{\text{CP}}$}
\newcommand{\esb}{\textsc{ESSnuSB}}
\newcommand{\hk}{\textsc{Hyper-Kamiokande}}
\newcommand{\gn}{\textsc{Genie}}
\newcommand{\ws}{\textsc{WCSim}}
\newcommand{\fq}{\textsc{fiTQun}}
\newcommand{\gf}{\textsc{Geant4}}

\newcommand{\NLLmath}{\tit{NLL}}

\newcommand{\NLL}{$\NLLmath$}

\title{The ESS neutrino super-beam near detector}

\author*[a,1]{Alexander Burgman}
\author[a,2]{Joochun Park}
\author[a]{Joakim Cederkäll}
\author[a]{Peter Christiansen}

\affiliation[a]{Department of Physics, Lund University,\\
  P.O Box 118, 221 00 Lund, Sweden}
\note{For the {\esb} Collaboration}
\note{Now at the Center for Exotic Nuclear Studies, Institute for Basic Science, 34126 Daejeon, Korea}

\emailAdd{alexander.burgman@nuclear.lu.se}

\abstract{
The ESS Neutrino Super-Beam ({\esb}) is a proposed long-baseline neutrino oscillation experiment,
performed with a high-intensity neutrino beam, to be developed as an extension to
the European Spallation Source proton linac currently under construction in Lund, Sweden.
The neutrinos would be detected with the near and far detectors of the experiment,
the former within several hundred meters of the neutrino production point and the latter within several hundred kilometers.
The far detector will consist of a megaton-scale water-Cherenkov detector,
and the near detector will consist of a kiloton-scale water-Cherenkov detector in combination with
a fine-grained tracking detector and an emulsion detector.
The purpose of the near detector is to constrain the flux of the neutrino beam as well as to extract
the electron-neutrino interaction cross-section in water, which requires high-performance
energy reconstruction and particle flavor identification techniques.
These measurements are crucial for the neutrino oscillation measurements
that will be conducted using the far detector.

Year 2021 sees the finalization of the conceptual design of the near detector after a thorough evaluation
of the performance of a number of different design options,
and a characterization of the neutrino reconstruction and flavor identification performances.
In this talk we report on these studies.
}

\FullConference{%
  *** The European Physical Society Conference on High Energy Physics (EPS-HEP2021), ***\\
  *** 26-30 July 2021 ***\\
  *** Online conference, jointly organized by Universität Hamburg and the research center DESY ***
}

\DeclareGraphicsExtensions{.pdf,.png,.jpg}
\graphicspath{ {./figs/} }
\begin{document}
\maketitle


\section{Introduction}

A possible explanation for the apparent asymmetry in the matter-antimatter abundances in the Universe is a simultaneous violation of the Charge and Parity symmetries (CP) in the leptonic sector. 
Recent measurements by the 
T2K experiment, making use of the Super-Kamiokande 
detector, indicate a non-zero value of the leptonic CP violation, quantified by the CP violating phase {\dcp}~\cite{T2K:2019bcf}.
In order to precisely measure the leptonic CP violation a new generation of long baseline water-Cherenkov detector -based neutrino experiments is proposed,
including the European Spallation Source neutrino Super Beam ({\esb})~\cite{ESSnuSB:2013dql,Wildner:2015yaa} and the Hyper-Kamiokande 
experiment~\cite{Hyper-Kamiokande:2016srs}.
These experiments will also be sensitive to explore additional open questions in modern fundamental physics, such as the ordering of the neutrino mass eigenstates, proton decay, and the detection of various astrophysically produced neutrinos.


\section{The {\esb} experiment and near detector}

The ESS neutrino super-beam experiment is a proposed long-baseline neutrino oscillation experiment, conducted using a high-intensity beam of neutrinos produced at the European Spallation Source (ESS) in Lund, Sweden, and detected in a water-Cherenkov detector \SI{360}{}--\SI{540}{\km} downstream of the source. 
The purpose of the {\esb} experiment is to measure the oscillation rates of muon-neutrinos into electron-neutrinos as well as muon-antineutrinos into electron-antineutrinos, and subsequently derive the value of the leptonic CP-violation parameter, {\dcp}.

In order to produce the high-intensity neutrino beam, several adjustments to the ESS facility are required, including an upgrade of the ESS proton linac to achieve the dedicated delivery of a \SI{5}{\MW} intensity $H^-$ ion beam with \SI{2.5}{\GeV} kinetic energy and \SI{14}{\Hz} repetition rate.
The ions would be stripped of all electrons and injected into a dedicated accumulator ring where the pulses are compressed to \SI{1.1}{\micro\s} in length. 
Four target stations receive the compressed proton bunches for collisions with a titanium bead target, thus producing a particle shower predominantly populated by charged pions.
The pions are sign-selected and collimated using magnetic horns, and directed into the decay tunnel where they decay to muons and muon-neutrinos.
The muons are then stopped by the beam-dump at the end of the decay tunnel, while the neutrino beam propagates towards the detectors.
A small population of electron-neutrinos is also produced from muon decays into electrons.
Selecting positive (negative) pions produces a beam of muon-(anti)neutrinos.

The majority of the beam neutrinos have energies between \SI{200}{} and \SI{700}{\MeV}, with maximum intensity at \SI{400}{\MeV}. The beam first arrives at the near detector (ND), \SI{250}{\m} from the production point, and later at the far detector, \SI{360}{}--\SI{540}{\km} away.
The far detector is currently modeled as a \SI{540}{\kt} water-Cherenkov detector situated at a depth of \SI{1}{\km} in the Swedish bedrock.

The purposes of the near detector are to measure the total neutrino flux of $\sim\SI{e7}{}$ events per running-year (\SI{200}{\d}, equivalent to \SI{2.16e23}{} protons-on-target), and to measure the interaction cross-section for electron-(anti)neutrinos incident on nuclei in water, $\sigma_{\nu_eN}$.
The latter is essential for the appearance-measurement of electron-(anti)neutrinos in the far detector. 
The ND will consist of two main components: a fine-grained plastic-scintillator tracking detector and a cylindrical water-Cherenkov detector.
The tracking detector is based on the Hyper-Kamiokande Super-FGD~\cite{Blondel:2017orl}, and will consist of \SI{e6}{} plastic scintillator cubes arranged in a $(1.4\times1.4\times0.5)~\SI{}{\m\cubed}$ block.
The water-Cherenkov detector tank will have a length of \SI{11}{\m} and a radius of \SI{4.7}{\m}, and have its central axis aligned with the neutrino beam direction.
The inner-surface fiducial coverage will be \SI{30}{\percent} using \SI{3.5}{\inch} diameter PMTs.
The possibility of including a third component, based on the nuclear emulsion detector used in the \textsc{Ninja} experiment~\cite{Odagawa:2020kqr}, is under investigation.

Negligible oscillation occurs before the neutrino beam arrives at the ND, yielding a beam of $>\SI{99.5}{\percent}$ muon-neutrinos with the remainder dominated by electron-neutrinos.
The low fraction of $\nu_e$ presents a challenge for measuring $\sigma_{\nu_eN}$, which requires an efficient $\nu_e$ event selection strategy.


\hide{
\begin{figure}[ht]
\begin{tikzpicture}
    \fill[yellow] (-0.5\textwidth,-0.1) rectangle (0.5\textwidth,0.1);
    \node[align=center] at (0.0,0.0) {\includegraphics[width=8.0cm]{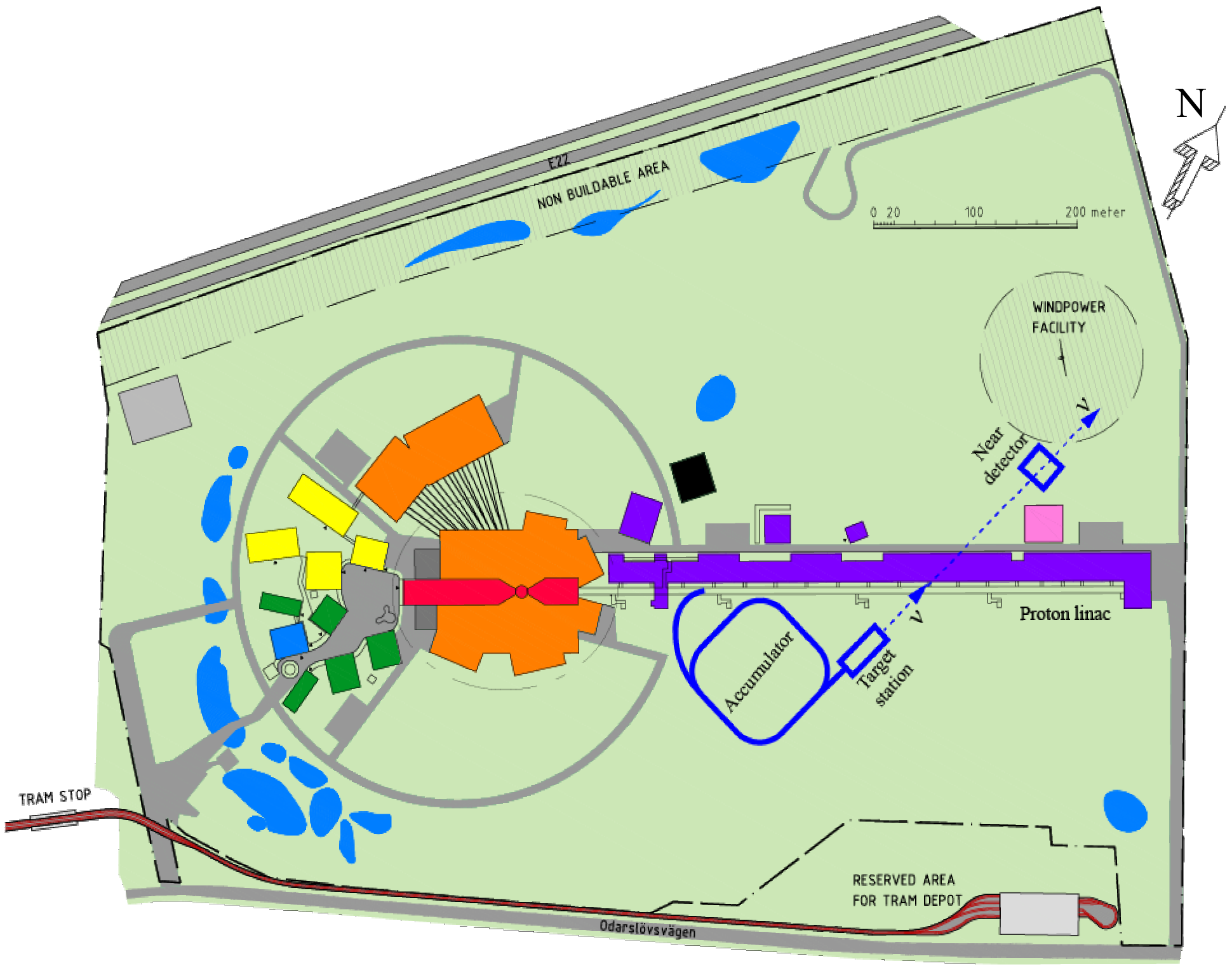}};
\end{tikzpicture}
\caption{the essnusb layout
\label{fig:essnusb_layout}}
\end{figure}
}

\hide{
\begin{figure}[t]
\begin{tikzpicture}
    \fill[yellow] (-0.5\textwidth,-0.1) rectangle (0.5\textwidth,0.1);
    \node[align=center] at (-4,-0.8) {\includegraphics[width=4.0cm]{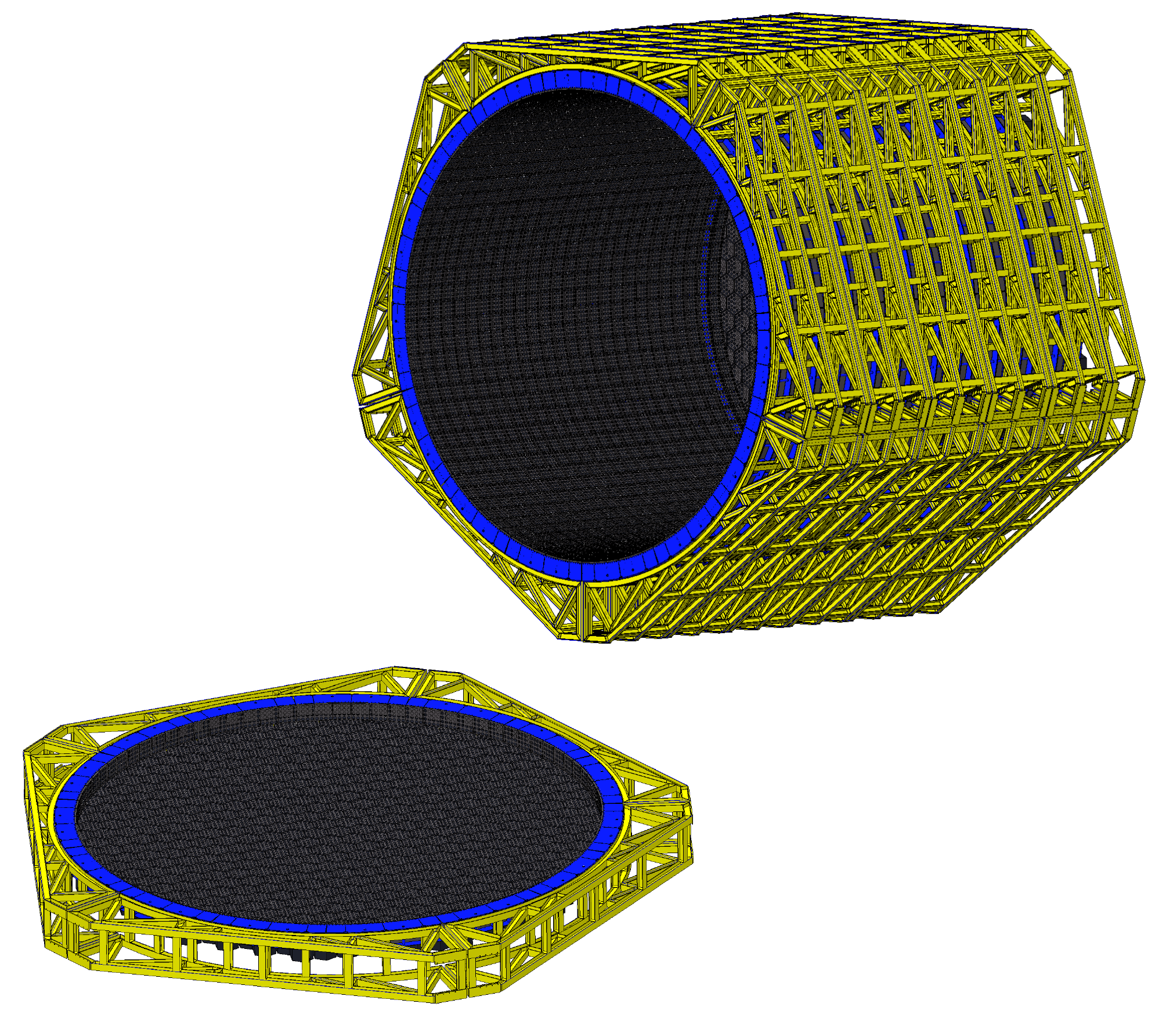}};
    \node[align=center] at (2.5,0.0) {\includegraphics[width=8.0cm]{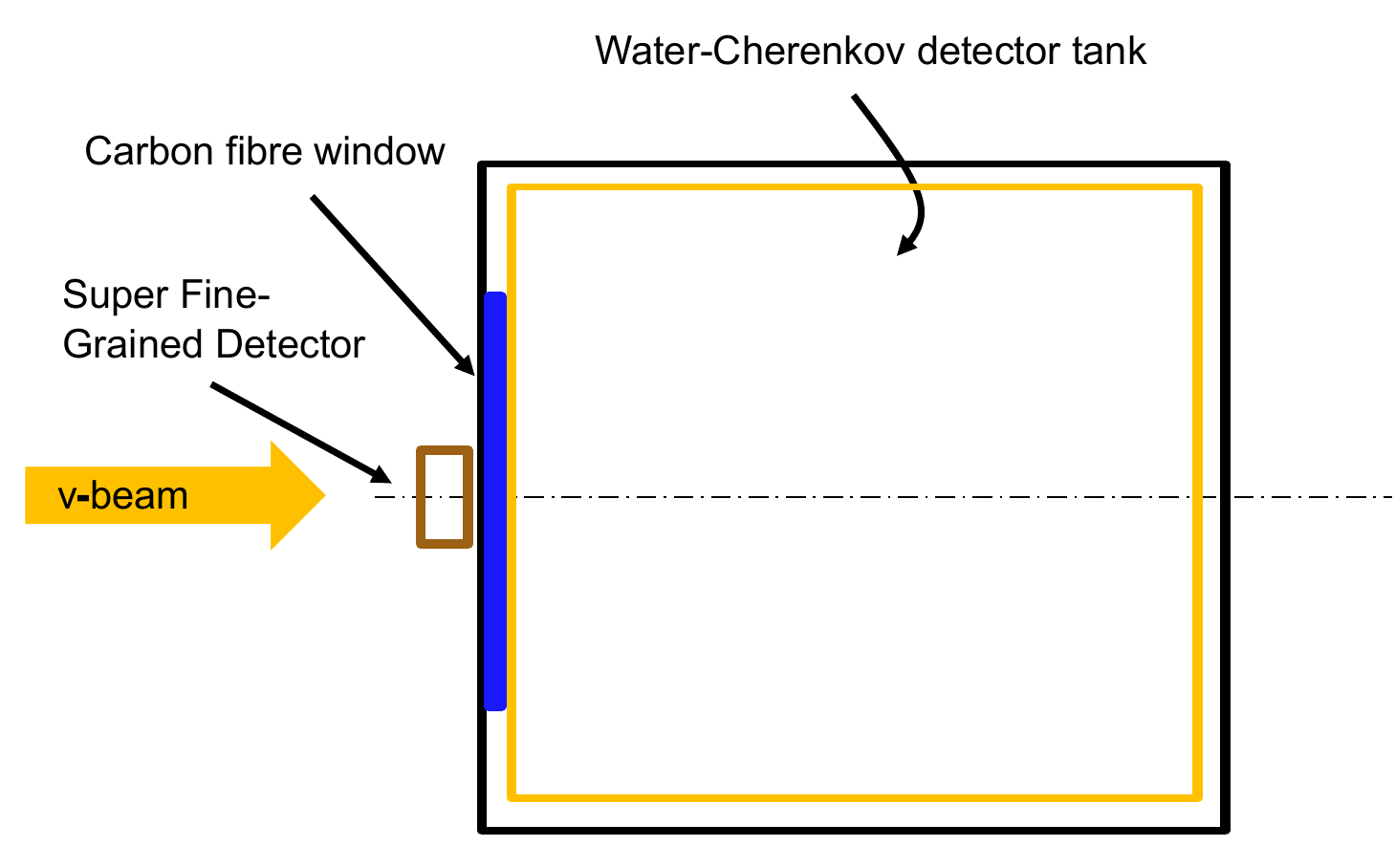}};
\end{tikzpicture}
\caption{the nd design
\label{fig:nd_design}}
\end{figure}
}


\section{Electron-neutrino selection}

In order to develop a reliable scheme for electron-neutrino event selection a two-step process was adopted:
(1) distinguishing electron events from muon events, and
(2) distinguishing electron-neutrino events from muon-neutrino events.


\subsection{Distinguishing electrons from muons}

In the first step, samples of electron and muon events were simulated in a model of the water-Cherenkov component of the ND using the {\ws} software~\cite{wcsimgithub}, a {\gf} based~\cite{GEANT4:2002zbu,Allison:2006ve,Allison:2016lfl} water-Cherenkov detector simulator.
The simulated charged-particle events were homogeneously and isotropically distributed in the detector tank, and uniformly distributed in kinetic energy up to \SI{1}{\GeV}.
The events were reconstructed using the {\fq} software~\cite{fitqun:2017,Super-Kamiokande:2019gzr}, developed by the {\hk} collaboration.%
\hide{
The {\fq} software applies several particle hypotheses to each event, and for each hypothesis it returns the most likely vertex position and direction, the particle energy, along with the negative-log-likelihood ({\NLL}) associated with the hypothesis.
{\fq} also supports modeling multiple sub-events, where it applies each particle hypothesis to each individual sub-event.
The {\NLL} value is lower for more likely hypotheses, and for the present analysis events are identified as electron candidate events if their electron and muon hypotheses {\NLL}, ($\NLLmath^e$ and $\NLLmath^\mu$, respectively) satisfy the following relation.
\begin{align}
    \frac{\NLLmath^\mu}{\NLLmath^e}\geq1.01
    \label{eqn:nll_eid}
\end{align}

After applying {\fq} to the samples the samples were studied in several reconstructed variables.
}
Several selection criteria were determined to reduce the number of misreconstructed and misidentified events, with an emphasis on the muon events that were mistakenly identified as electron events.
\begin{description}
    \item[Sub-Cherenkov cut] Rejecting muon events below the Cherenkov threshold\\[-7mm]
    \item[Reconstruction quality cut] Rejecting events with poor reconstruction conditions, indicated by a low registered brightness and a reconstructed vertex position close to the detector tank wall\\[-7mm]
    \item[Cherenkov-ring resolution cut] Rejecting events that are too close to the tank wall in the reconstructed propagation direction of the lepton\\[-7mm]
    \item[Vertex-reconstruction discrepancy cut] Rejecting events with a large difference in vertex position and direction between the electron and muon reconstruction hypotheses
\end{description}

The final selection efficiency is \SI{46.3}{\percent} for electron events, with a \SI{1.6}{\percent} misidentification rate, and \SI{43.3}{\percent} for muon events with a \SI{0.3}{\percent} misidentification rate.

\subsection{Distinguishing electron-neutrinos from muon-neutrinos}

In the second step a sample of neutrino interaction vertices were simulated using the {\gn} neutrino interaction generator~\cite{Andreopoulos:2009rq,Andreopoulos:2015wxa,Tena-Vidal:2021rpu}, and were weighted to the unoscillated expected neutrino flux.
The neutrino vertices were homogeneously distributed in the detector tank, using the {\ws} software, with the neutrino direction aligned with the neutrino beam axis.
The numbers of expected neutrino interactions in the detector per \SI{200}{\d} running-year are shown in Table~\ref{tbl:nev_nu} along with the expected number of triggering events.
The events were again reconstructed using the {\fq} software, and the selection criteria developed in the previous step were applied.

Two additional criteria were developed to address event-features that are relevant in the selection of electron-neutrino events, but inaccessible in pure charged-lepton events.
\begin{description}
    \item[Pion-like cut] Rejecting events that are identified as electrons but have a high pion likelihood\\[-7mm]
    \item[Multi-subevent cut] Rejecting electron-like events with multiple identified subevents
\end{description}

\hide{
\begin{figure}[t]
\begin{tikzpicture}
    \fill[white] (-0.5\textwidth,-0.1) rectangle (0.5\textwidth,0.1);
    \node[anchor=center] at (0,0            ) { \includegraphics[width=8cm]{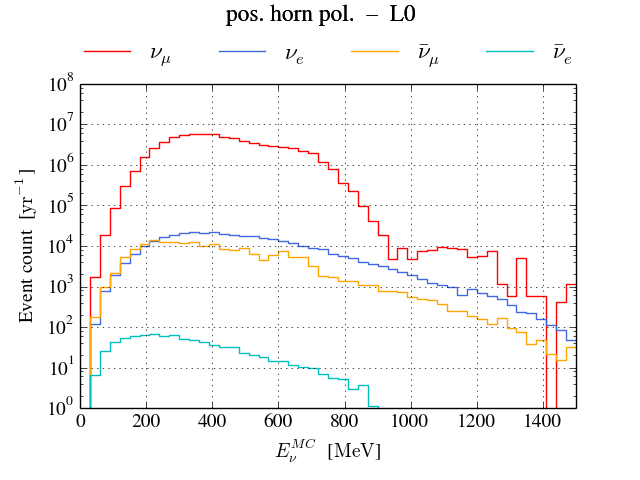} };
    \fill[white] (-0.12\textwidth,0.17\textwidth) rectangle (0.12\textwidth,0.20\textwidth);
\end{tikzpicture}
\caption{Unoscillated expected energy spectra of the neutrino flux.
\label{fig:spectrum_energy}}
\end{figure}
}

The number of electron-identified events that remain at each selection level are shown in Table~\ref{tbl:nev_nu_eid} for the four considered neutrino species.
This event selection reduces the muon-neutrino event rate in the electron-like sample to a similar level as the electron-neutrino rate, for each focusing horn polarity.
For the positive and negative horn polarities the signal-to-background ratios are 0.73 and 0.97, respectively. 

\begin{table}[b]
\centering
\begin{tabular}{ r r r r r | l }
    \hline
    Selection level & $\nu_\mu$ & $\nu_e$ & $\bar{\nu}_\mu$ & $\bar{\nu}_e$ & Horn polarization \\
    \hline
    Total interactions & \SI{7.25e+07}{} & \SI{3.57e+05}{} & \SI{1.89e+05}{} & \SI{833}{} & Positive \\
    Trigger & \SI{3.81e+07}{} & \SI{5.61e+04}{} & \SI{9.09e+04}{} & \SI{93.5}{} \\
    \hline
    Total interactions & \SI{6.88e+05}{} & \SI{4740}{} & \SI{1.39e+07}{} & \SI{4.12e+04}{} & Negative \\
    Trigger & \SI{3.48e+05}{} & \SI{645}{} & \SI{6.84e+06}{} & \SI{5040}{} \\
    \hline
\end{tabular}
\caption{The number of expected neutrino interactions for each neutrino flavor as well as the expected number of event triggers. Shown for both positive and negative polarization of the focusing horn.
\label{tbl:nev_nu}}
\end{table}

\begin{table}[b]
\centering
\begin{tabular}{ r r r r r | l }
    \hline
    Selection level & $\nu_\mu$ ($e^\text{ID}$) & $\nu_e$ ($e^\text{ID}$) & $\bar{\nu}_\mu$ ($e^\text{ID}$) & $\bar{\nu}_e$ ($e^\text{ID}$) & Horn polarization \\
    \hline
    Trigger & \SI{1.09e+07}{} & \SI{5.26e+04}{} & \SI{2.66e+04}{} & \SI{88.2}{} & Positive \\
    Charged-lepton cuts & \SI{5.72e+05}{} & \SI{2.29e+04}{} & \SI{1430}{} & \SI{35.8}{} & ~ \\
    Neutrino cuts & \SI{1.50e+04}{} & \SI{1.10e+04}{} & \SI{41.1}{} & \SI{32.7}{} & ~ \\
    \hline
    Trigger & \SI{1.08e+05}{} & \SI{605}{} & \SI{1.87e+06}{} & \SI{4740}{} & Negative \\
    Charged-lepton cuts & \SI{6720}{} & \SI{259}{} & \SI{5.12e+04}{} & \SI{2120}{} & ~ \\
    Neutrino cuts & \SI{123}{} & \SI{123}{} & \SI{1930}{} & \SI{1860}{} & ~ \\
    \hline
\end{tabular}
\caption{The number of electron-identified events ($e^\text{ID}$) at each selection level for the four considered neutrino species. Shown for both positive and negative polarization of the focusing horn.
\label{tbl:nev_nu_eid}}
\end{table}

In order to properly measure the interaction cross-section for electron-neutrinos incident on nucleons in water, $\sigma_{\nu_eN}$, the neutrino energy must be reliably calculated.
Assuming a quasi-elastic charged-current interaction, the neutrino energy $E_\nu$ is calculated using the following formula:
\begin{align}
    E_\nu = \frac{m_F^2-m_\tit{IB}^2-m_l^2+2m_BE_l}{2\left(m_B-E_l+p_l\cos\theta_l\right)}
    \label{eqn:energy_nu}
\end{align}

Here, $m_F$ represents the final state mass of the nucleon, $m_\tit{IB}=m_I-E_B$ represents the bound state energy of the target nucleon, with $m_I$ as the initial state free nucleon mass and $E_B$ as the binding energy.
For the purposes of this analysis, the nucleon mass is assumed as the proton mass, and $E_B$ is approximated as the $^{16}\text{O}$ binding energy of \SI{27}{\MeV}.
The subscript $l$ represents the final state charged lepton, and thus $m_l$, $E_l$, $p_l$ and $\theta_l$ respectively represent its mass, energy, absolute momentum and angle relative the neutrino beam axis.

\begin{figure}[t]
\begin{tikzpicture}
    \fill[white] (-0.5\textwidth,-0.1) rectangle (0.5\textwidth,0.1);
    \node[anchor=center] at (-0.25\textwidth,0) { \includegraphics[width=8cm]{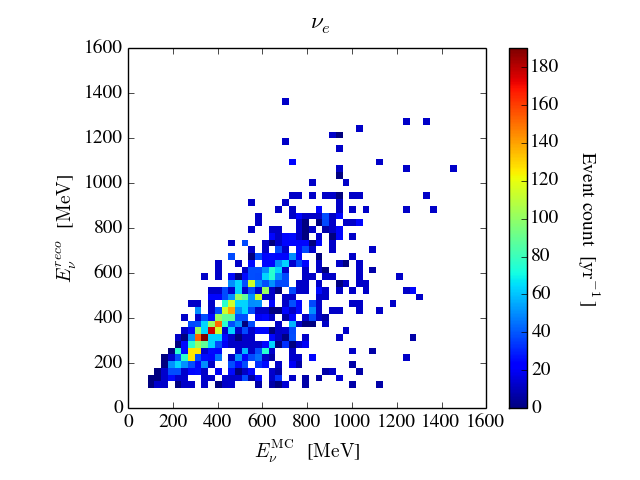} };
    \node[anchor=center] at ( 0.25\textwidth,0) { \includegraphics[width=8cm]{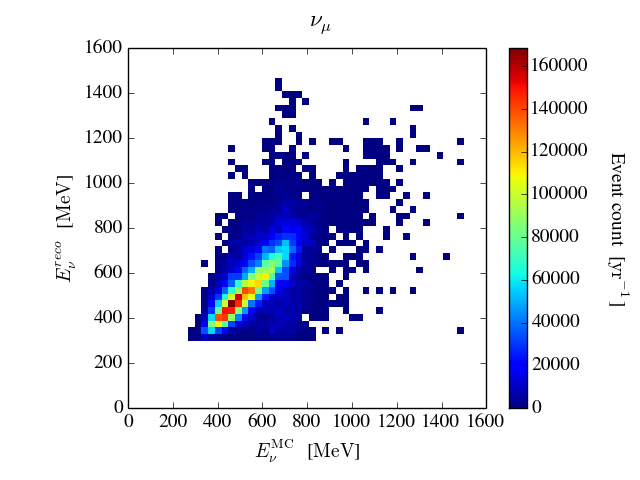} };
    \node[anchor=west] at (-0.38\textwidth, 1.5) { \color{blue3}\huge$\nu_e$  };
    \node[anchor=west] at ( 0.12\textwidth, 1.5) { \color{red3}\huge$\nu_\mu$ };
    \fill[white] (-0.3\textwidth,2.5) rectangle (-0.2\textwidth,2.9);
    \fill[white] (0.3\textwidth,2.5) rectangle (0.2\textwidth,2.9);
\end{tikzpicture}
\caption{Event distributions over the true and reconstructed neutrino energies per running-year with positive horn polarity, after the full event selection.
\textit{Left:} electron-neutrino events.
\textit{Right:} muon-neutrino events.
\label{fig:reco_e_nu}}
\end{figure}
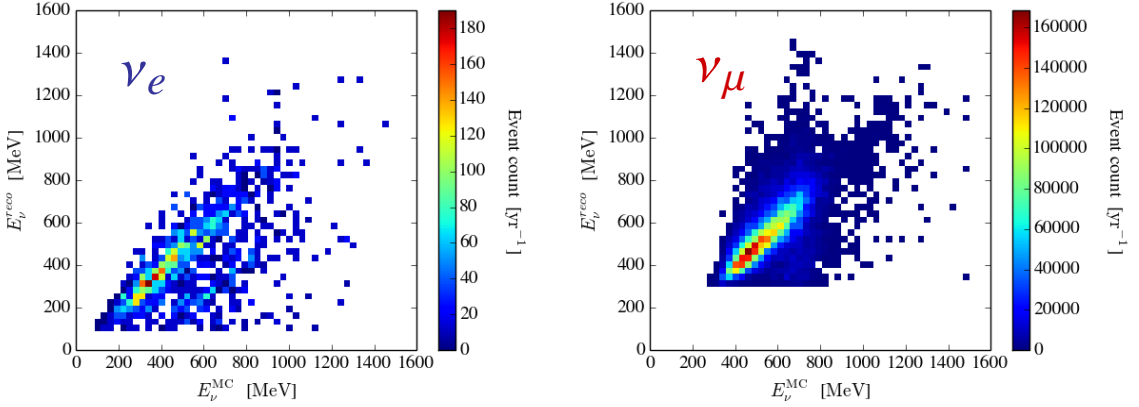

The electron- and muon-neutrino event distributions are shown in Figure~\ref{fig:reco_e_nu} over the true and reconstructed neutrino energies, $E_\nu^\text{MC}$ and $E_\nu^{\tit{reco}}$ respectively, after the full event selection.
The difference between the true and reconstructed energies is distributed with root-mean-square values of \SI{115}{\MeV} (\SI{101}{\MeV}) for muon-neutrinos and \SI{212}{\MeV} (\SI{138}{\MeV}) for electron-neutrinos for positive (negative) horn polarity.

\section{Summary}

The {\esb} long-baseline experiment is proposed to measure the leptonic CP violation with unprecedented precision through the measurement of neutrino oscillations in a high-intensity and low-energy neutrino beam.
The neutrinos will be detected by interactions in the megaton-scale water-Cherenkov far detector, as well as the near detector system. 
The kiloton-scale water-Cherenkov near detector will be an essential tool for measuring the electron-neutrino interaction cross-section with nucleons, for which the low $\nu_e$ fraction in the neutrino beam requires a dedicated event selection.

In this talk we have presented the current $\nu_e$ event selection strategy,
where the electron-neutrino contribution was enhanced from $<\SI{2}{\permille}$
at the trigger level to $\sim\SI{50}{\percent}$
at the final selection level.
This selection still accepts a substantial fraction of triggered electron-neutrino events, \SI{20}{\percent} (\SI{35}{\percent}) for positive (negative) horn polarity, and demonstrates a good performance for neutrino energy reconstruction.
This will enable the measurement of the neutrino-nucleon interaction cross-section.


\acknowledgments

This project received funding from the European Union’s Horizon 2020 research and innovation programme under grant agreement No.\ 777419.
Thanks are extended to C.~Vilela, E.~O’Sullivan, H.~Tanaka, B.~Quilain and M.~Wilking for their assistance with using the {\ws} and {\fq} software packages.


\bibliography{bibl}

\providecommand{\href}[2]{#2}\begingroup\raggedright\begin{thebibliography}{10}

\bibitem{T2K:2019bcf}
{\scshape T2K} collaboration, \emph{{Constraint on the
  matter\textendash{}antimatter symmetry-violating phase in neutrino
  oscillations}},
  \href{https://doi.org/10.1038/s41586-020-2177-0}{\emph{Nature} {\bfseries
  580} (2020) 339} [\href{https://arxiv.org/abs/1910.03887}{{\ttfamily
  1910.03887}}].

\bibitem{ESSnuSB:2013dql}
{\scshape ESSnuSB} collaboration, \emph{{A very intense neutrino super beam
  experiment for leptonic CP violation discovery based on the European
  spallation source linac}},
  \href{https://doi.org/10.1016/j.nuclphysb.2014.05.016}{\emph{Nucl. Phys. B}
  {\bfseries 885} (2014) 127}
  [\href{https://arxiv.org/abs/1309.7022}{{\ttfamily 1309.7022}}].

\bibitem{Wildner:2015yaa}
E.~Wildner et~al., \emph{{The Opportunity Offered by the \textsc{ESSnuSB}
  Project to Exploit the Larger Leptonic CP Violation Signal at the Second
  Oscillation Maximum and the Requirements of This Project on the ESS
  Accelerator Complex}}, \href{https://doi.org/10.1155/2016/8640493}{\emph{Adv.
  High Energy Phys.} {\bfseries 2016} (2016) 8640493}
  [\href{https://arxiv.org/abs/1510.00493}{{\ttfamily 1510.00493}}].

\bibitem{Hyper-Kamiokande:2016srs}
{\scshape Hyper-Kamiokande} collaboration, \emph{{Physics potentials with the
  second \textsc{Hyper-Kamiokande} detector in Korea}},
  \href{https://doi.org/10.1093/ptep/pty044}{\emph{PTEP} {\bfseries 2018}
  (2018) 063C01} [\href{https://arxiv.org/abs/1611.06118}{{\ttfamily
  1611.06118}}].

\bibitem{Blondel:2017orl}
A.~Blondel et~al., \emph{{A fully active fine grained detector with three
  readout views}},
  \href{https://doi.org/10.1088/1748-0221/13/02/P02006}{\emph{JINST} {\bfseries
  13} (2018) P02006} [\href{https://arxiv.org/abs/1707.01785}{{\ttfamily
  1707.01785}}].

\bibitem{Odagawa:2020kqr}
{\scshape Ninja} collaboration, \emph{{Prospect and status of the physics run
  of the \textsc{Ninja} experiment}},
  \href{https://doi.org/10.22323/1.369.0144}{\emph{PoS} {\bfseries NuFact2019}
  (2020) 144}.

\bibitem{wcsimgithub}
``\ws~github.'' \url{https://github.com/WCSim/WCSim}.

\bibitem{GEANT4:2002zbu}
{\scshape Geant4} collaboration, \emph{{\textsc{Geant4}--a simulation
  toolkit}}, \href{https://doi.org/10.1016/S0168-9002(03)01368-8}{\emph{Nucl.
  Instrum. Meth. A} {\bfseries 506} (2003) 250}.

\bibitem{Allison:2006ve}
J.~Allison et~al., \emph{{\textsc{Geant4} developments and applications}},
  \href{https://doi.org/10.1109/TNS.2006.869826}{\emph{IEEE Trans. Nucl. Sci.}
  {\bfseries 53} (2006) 270}.

\bibitem{Allison:2016lfl}
J.~Allison et~al., \emph{{Recent developments in \textsc{Geant4}}},
  \href{https://doi.org/10.1016/j.nima.2016.06.125}{\emph{Nucl. Instrum. Meth.
  A} {\bfseries 835} (2016) 186}.

\bibitem{fitqun:2017}
{\scshape T2K} collaboration, \emph{{Improving the T2K Oscillation Analysis
  With \textsc{fiTQun}: A New Maximum-Likelihood Event Reconstruction for
  \textsc{Super-Kamiokande}}},
  \href{https://doi.org/10.1088/1742-6596/888/1/012066}{\emph{J. Phys. Conf.
  Ser.} {\bfseries 888} (2017) }.

\bibitem{Super-Kamiokande:2019gzr}
{\scshape Super-Kamiokande} collaboration, \emph{{Atmospheric Neutrino
  Oscillation Analysis with Improved Event Reconstruction in
  \textsc{Super-Kamiokande} IV}},
  \href{https://doi.org/10.1093/ptep/ptz015}{\emph{PTEP} {\bfseries 2019}
  (2019) 053F01} [\href{https://arxiv.org/abs/1901.03230}{{\ttfamily
  1901.03230}}].

\bibitem{Andreopoulos:2009rq}
C.~Andreopoulos et~al., \emph{{The \textsc{Genie} Neutrino Monte Carlo
  Generator}}, \href{https://doi.org/10.1016/j.nima.2009.12.009}{\emph{Nucl.
  Instrum. Meth. A} {\bfseries 614} (2010) 87}
  [\href{https://arxiv.org/abs/0905.2517}{{\ttfamily 0905.2517}}].

\bibitem{Andreopoulos:2015wxa}
C.~Andreopoulos, C.~Barry, S.~Dytman, H.~Gallagher, T.~Golan, R.~Hatcher
  et~al., \emph{{The \textsc{Genie} Neutrino Monte Carlo Generator: Physics and
  User Manual}},  \href{https://arxiv.org/abs/1510.05494}{{\ttfamily
  1510.05494}}.

\bibitem{Tena-Vidal:2021rpu}
{\scshape Genie} collaboration, \emph{{Neutrino-Nucleon Cross-Section Model
  Tuning in \textsc{Genie} v3}},
  \href{https://arxiv.org/abs/2104.09179}{{\ttfamily 2104.09179}}.

\end{thebibliography}\endgroup


\end{document}